\documentclass[12pt,thmsa]{article}
\def \eq {\begin{equation}}
\def \fim-eq {\end{equation}}
\begin{document}

\author{E. S. Guerra \\
%EndAName
Departamento de F\'{\i}sica \\
Universidade Federal Rural do Rio de Janeiro \\
Cx. Postal 23851, 23890-000 Serop\'edica, RJ, Brazil \\
email: emerson@ufrrj.br\\
}
\title{ TELEPORTATION OF ATOMIC STATES VIA CAVITY QED FOR A CAVITY PREPARED
IN A SUPERPOSITION OF ZERO AND ONE FOCK STATES}
\maketitle

\begin{abstract}
\noindent In this article we discuss two schemes of teleportation of atomic
states. In the first scheme we consider atoms in a three-level cascade
configuration and in the second scheme we consider atoms in a three-level
lambda configuration. The experimental realization proposed makes use of
cavity Quatum Electrodynamics involving the interaction of Rydberg atoms
with a micromaser cavity prepared in a state $|\psi \rangle _{C}=(|0\rangle
+|1\rangle )/\sqrt{2}$.

\ \newline

\end{abstract}

\section{\protect\bigskip INTRODUCTION\protect\bigskip}

Quantum information and quantum computation are important and active fields
of research \cite{Nielsen, MathQC, PhysQI}. Teleportation, proposed by
Bennett \textit{et al }\cite{Bennett}, has important applications in quantum
information and quantum computation \cite{Nielsen}. There has been a lot of
theoretical proposals of schemes of teleportation. In Ref.\ \cite{DavZag} it
is proposed a teleportation scheme based on cavity QED for the teleportation
of an atomic state where the cavities are prepared in a entangled state of
zero and one Fock states. In Ref. \cite{GuerraTelCascatCS} it is proposed a
scheme to teleport an atomic state for atoms in a cascade configuration
making use of cavities prepared in a coherent state. In Ref. \cite%
{GuerraTelLambatCS} it is proposed a scheme to teleport an atomic state for
atoms in a lambda configuration making use of cavities prepared in a
coherent state. In Ref. \cite{SongLuZhangGuo} it is presented a scheme of
teleportation where a superposition of zero and one Fock states is
teleported via cavity QED using atoms in a lambda configuration. An
interesting proposition of generating EPR states and realization of
teleportation using a dispersive atom-field interaction is presented in \cite%
{ZengGuo}. Teleportation has already been realized experimentally \cite%
{ExpTel, PhysQI}. The superposition principle together with entanglement and
its consequence non-locality are the main ingredients in teleportation. The
goal in teleportation is to reproduce an unknown quantum state of a given
system in another system far apart of the original system. That is if Alice
has a system prepared in an unknown state, by teleportation, she is able to
transfer this state to Bob%
%TCIMACRO{\U{b4}}%
%BeginExpansion
\'{}%
%EndExpansion
s system. In order to do this, Alice and Bob share a Bell state\ \cite%
{PhysQI, Nielsen}\ (or EPR state \cite{EPR}) in which half of the Bell pair
is with Alice and the other half is with Bob. Then Alice and Bob perform a
certain prescription and communicate classically with each other. In the end
of the process Bob gets a state identical to the state of the original state
in which Alice system was prepared and the state of Alice%
%TCIMACRO{\U{b4}}%
%BeginExpansion
\'{}%
%EndExpansion
s system is destroyed since according to the no-cloning theorem \cite%
{Nocloning, Nielsen} it is not possible to clone a quantum state.

In this article we consider Rydberg atoms \cite{Rydat} interacting with a
superconducting cavity \cite{Haroche, Walther} prepared in a superposition
of a zero and a one Fock states. We develop two schemes of teleportation. In
the first scheme we consider atoms in a three-level cascade configuration
and in the second scheme we consider atoms in a three-level lambda
configuration.

\section{TELEPORTATION}

\bigskip

\subsection{\protect\bigskip ATOMS IN A CASCADE CONFIGURATION\protect\bigskip%
}

First let us present a scheme to prepare a Bell states. We start assuming
that we have a cavity $C$ prepared in the state 
\begin{equation}
|\psi \rangle _{C}=\frac{(|0\rangle +|1\rangle )}{\sqrt{2}}.  \label{PsiC}
\end{equation}%
In order to prepare this state, we send a two-level atom $A0$, with $%
|f_{0}\rangle $ and $|e_{0}\rangle $ being the lower and upper level
respectively, through a Ramsey cavity $R0$ in the lower state $\mid
f_{0}\rangle $ where the atomic states are rotated according to%
\begin{equation}
R_{0}=\frac{1}{\sqrt{2}}\left[ 
\begin{array}{cc}
1 & i \\ 
i & 1%
\end{array}%
\right] ,
\end{equation}%
that is,%
\begin{equation}
\mid f_{0}\rangle \rightarrow \frac{1}{\sqrt{2}}(i\mid e_{0}\rangle +\mid
f_{0}\rangle ),
\end{equation}%
and through $C$, for $A0$ resonant with the cavity. Under the
Jaynes-Cummings dynamics \cite{Orszag} (see\ (\ref{UJC}), for $\Delta =0$
and $gt=\pi /2$) we know that the state $|f_{0}\rangle |0\rangle $ does not
evolve, however, the state $|e_{0}\rangle |0\rangle $ evolves to $%
-i|f_{0}\rangle |1\rangle $. Then, for the cavity initially in the vacuum
state $|0\rangle $, we have%
\begin{equation}
\frac{(|f_{0}\rangle +i|e_{0}\rangle )}{\sqrt{2}}|0\rangle \longrightarrow
|f_{0}\rangle \frac{(|0\rangle +|1\rangle )}{\sqrt{2}}=|f_{0}\rangle |\psi
\rangle _{C}.
\end{equation}

Now let us consider a three-level cascade atom \ $Ak$ with $\mid
e_{k}\rangle ,\mid f_{k}\rangle $ and $\mid g_{k}\rangle $ being the upper,
intermediate and lower atomic state (see Fig. 1). We assume that the
transition $\mid f_{k}\rangle \rightleftharpoons \mid e_{k}\rangle $ is far
enough from resonance with the cavity central frequency such that only
virtual transitions occur between these states (only these states interact
with field in cavity $C$). In addition we assume that the transition $\mid
e_{k}\rangle \rightleftharpoons \mid g_{k}\rangle $ is highly detuned from
the cavity frequency so that there will be no coupling with the cavity
field. Here we are going to consider the effect of the atom-field
interaction taking into account only levels $\mid f_{k}\rangle $ and $\mid
g_{k}\rangle .$ We do not consider level $\mid e_{k}\rangle $ since it will
not play any role in our scheme. Therefore, we have effectively a two-level
system involving states $\mid f_{k}\rangle $ and $|g_{k}\rangle $.
Considering levels $\mid f_{k}\rangle $ and $\mid g_{k}\rangle ,$ we can
write an effective time evolution operator (see ( \ref{Ud})) 
\begin{equation}
U_{k}(t)=e^{i\varphi a^{\dagger }a}\mid f_{k}\rangle \langle f_{k}\mid
+|g_{k}\rangle \langle g_{k}\mid ,  \label{UCasc}
\end{equation}%
where the second term above was put by hand just in order to take into
account the effect of level $\mid g_{k}\rangle $. In (\ref{UCasc}) $a$ $%
(a^{\dagger })$ is the annihilation (creation) operator for the field in
cavity $C$, $\varphi =g^{2}\tau /$ $\Delta $, \ $g$ is the coupling
constant, $\Delta =\omega _{e}-\omega _{f}-\omega $ is the detuning \ where
\ $\omega _{e}$ and $\omega _{f}$ \ are the frequencies of the upper and
intermediate levels respectively and $\omega $ is the cavity field frequency
and $\tau $ is the atom-field interaction time. Let us take $\varphi =\pi $.
\ Now, let us assume that we let atom $A1$ to interact with cavity $C$
prepared in the state (\ref{PsiC}). Let us assume that atom $A1$ is prepared
in a Ramsey cavity $R1$ in a coherent superposition according to the
rotation matrix%
\begin{equation}
R=\frac{1}{\sqrt{2}}\left[ 
\begin{array}{cc}
1 & 1 \\ 
-1 & 1%
\end{array}%
\right] ,  \label{Rot1}
\end{equation}%
and we have 
\begin{equation}
\mid \psi \rangle _{A1}=\frac{1}{\sqrt{2}}(\mid f_{1}\rangle +\mid
g_{1}\rangle ).
\end{equation}%
Taking into account (\ref{UCasc}), after atom $A1$ has passed through the
cavity prepared in state (\ref{PsiC}), we get 
\begin{equation}
\mid \psi \rangle _{A1-C}=\frac{1}{2}[(\mid f_{1}\rangle +\mid g_{1}\rangle
)|0\rangle +(-\mid f_{1}\rangle +\mid g_{1}\rangle )|1\rangle ],
\end{equation}%
Now, if atom $A1$ enters a second Ramsey cavity $R2$ where the atomic states
are rotated according to the rotation matrix (\ref{Rot1}), we have 
\begin{eqnarray}
\frac{1}{\sqrt{2}}( &\mid &f_{1}\rangle +\mid g_{1}\rangle )\rightarrow \mid
f_{1}\rangle ,  \nonumber \\
\frac{1}{\sqrt{2}}(- &\mid &f_{1}\rangle +\mid g_{1}\rangle )\rightarrow
\mid g_{1}\rangle ,
\end{eqnarray}%
and, therefore, 
\begin{equation}
\mid \psi \rangle _{A1-C}=\frac{1}{\sqrt{2}}[\mid f_{1}\rangle |0\rangle
+\mid g_{1}\rangle |1\rangle ],
\end{equation}%
Now, let us prepare a two-level atom $A2$ in the Ramsey cavity $R3$. If atom 
$A2$ is initially in the state $\mid g_{2}\rangle $, according to the
rotation matrix (\ref{Rot1}), we have%
\begin{equation}
\mid \psi \rangle _{A2}=\frac{1}{\sqrt{2}}(\mid f_{2}\rangle +\mid
g_{2}\rangle ),
\end{equation}%
and let us send this atom through cavity $C$, assuming that \ for atom $A2$,
as above for atom $A1$, the transition $\mid f_{2}\rangle \rightleftharpoons
\mid g_{2}\rangle $ is highly detuned from the cavity central frequency.
Taking into account (\ref{UCasc}), after the atom has passed through the
cavity we get 
\begin{equation}
\mid \psi \rangle _{A1-A2-C}=\frac{1}{2}[\mid f_{1}\rangle (\mid
f_{2}\rangle +\mid g_{2}\rangle )|0\rangle +\mid g_{1}\rangle (-\mid
f_{2}\rangle +\mid g_{2}\rangle )|1\rangle ],
\end{equation}%
Then, atom $A2$ enters a Ramsey cavity $R4$ where the atomic states are
rotated according to the rotation matrix (\ref{Rot1}), that is,%
\begin{eqnarray}
\frac{1}{\sqrt{2}}( &\mid &f_{2}\rangle +\mid g_{2}\rangle )\rightarrow \mid
f_{2}\rangle ,  \nonumber \\
\frac{1}{\sqrt{2}}(- &\mid &f_{2}\rangle +\mid g_{2}\rangle )\rightarrow
\mid g_{2}\rangle ,
\end{eqnarray}%
and we get%
\begin{equation}
\mid \psi \rangle _{A1-A2-C}=\frac{1}{\sqrt{2}}(\mid f_{1}\rangle \mid
f_{2}\rangle |0\rangle +\mid g_{1}\rangle \mid g_{2}\rangle |1\rangle ),
\end{equation}%
In order to disentangle the atomic states of the cavity field state we now
send a two-level atom $A3,$ resonant with the cavity, with $|f_{3}\rangle $
and $|e_{3}\rangle $ being the lower and upper level respectively, through $%
C $. If $A3$ is sent in the lower state $|f_{3}\rangle $, under the
Jaynes-Cummings dynamics (see\ (\ref{UJC}), for $\Delta =0$ and $gt=\pi /2$)
we know that the state $|f_{3}\rangle |0\rangle $ does not evolve, however,
the state $|f_{3}\rangle |1\rangle $ evolves to $-i|e_{3}\rangle |0\rangle $%
. Then we get%
\begin{equation}
\mid \psi \rangle _{A1-A2-C}=\frac{1}{\sqrt{2}}(\mid f_{1}\rangle \mid
f_{2}\rangle |f_{3}\rangle -i\mid g_{1}\rangle \mid g_{2}\rangle
|e_{3}\rangle )|0\rangle ,
\end{equation}%
Now we let atom $A3$ to enter a Ramsey cavity $R5$ where the atomic states
are rotated according to the rotation matrix%
\begin{equation}
R=\frac{1}{\sqrt{2}}\left[ 
\begin{array}{cc}
1 & i \\ 
i & 1%
\end{array}%
\right] ,  \label{Rot2}
\end{equation}%
that is,%
\begin{eqnarray}
&\mid &e_{3}\rangle \rightarrow \frac{1}{\sqrt{2}}(\mid e_{3}\rangle +i\mid
f_{3}\rangle ),  \nonumber \\
&\mid &f_{3}\rangle \rightarrow \frac{1}{\sqrt{2}}(i\mid e_{3}\rangle +\mid
f_{3}\rangle ),
\end{eqnarray}%
and we get 
\begin{equation}
\mid \psi \rangle _{A1-A2}=\frac{1}{2}[\mid f_{1}\rangle \mid f_{2}\rangle
(i\mid e_{3}\rangle +\mid f_{3}\rangle )-i\mid g_{1}\rangle \mid
g_{2}\rangle (\mid e_{3}\rangle +i\mid f_{3}\rangle )],
\end{equation}%
and if we detect atom $A3$ in state $|f_{3}\rangle $ finally we get the Bell
state 
\begin{equation}
\mid \Phi ^{+}\rangle _{A1-A2}=\frac{1}{\sqrt{2}}(\mid f_{1}\rangle \mid
f_{2}\rangle +\mid g_{1}\rangle \mid g_{2}\rangle ),  \label{PHI+Casc}
\end{equation}%
which is an entangled state of atoms $A1$ and $A2$. If we detect atom $A3$
in state $|e_{3}\rangle $ we get 
\begin{equation}
\mid \Phi ^{-}\rangle _{A1-A2}=\frac{1}{\sqrt{2}}(\mid f_{1}\rangle \mid
f_{2}\rangle -\mid g_{1}\rangle \mid g_{2}\rangle ),  \label{PHI-Casc}
\end{equation}%
Now, if we apply an extra rotation on the states of atom $A2$ in (\ref%
{PHI+Casc}) in a Ramsey cavity $R,$ according to the rotation matrix 
\begin{equation}
R=\mid f_{2}\rangle \langle g_{2}|+\mid g_{2}\rangle \langle f_{2}|,
\label{RC}
\end{equation}%
we get%
\begin{equation}
\mid \Psi ^{+}\rangle _{A1-A2}=\frac{1}{\sqrt{2}}(\mid f_{1}\rangle \mid
g_{2}\rangle +\mid g_{1}\rangle \mid f_{2}\rangle ),  \label{PSI+Casc}
\end{equation}%
and applying (\ref{RC}) to (\ref{PHI-Casc}) we get%
\begin{equation}
\mid \Psi ^{-}\rangle _{A1-A2}=\frac{1}{\sqrt{2}}(\mid f_{1}\rangle \mid
g_{2}\rangle -\mid g_{1}\rangle \mid f_{2}\rangle ).  \label{PSI-Casc}
\end{equation}%
The states (\ref{PHI+Casc}), (\ref{PHI-Casc}), (\ref{PSI+Casc}) and (\ref%
{PSI-Casc}) form a Bell basis \cite{Nielsen, PhysQI} which are a complete
orthonormal basis for the system $A1-A2$.

Now let us see how to distinguish the four states which form the Bell basis.
Let us assume we have a cavity $C$ prepared in the state (\ref{PsiC}).
Notice that if we send atoms $A1$ and $A2$ through $C$ in the state (\ref%
{PHI+Casc}) or (\ref{PHI-Casc}) we have%
\begin{equation}
\mid \Phi ^{\pm }\rangle _{A1-A2}\frac{(|0\rangle +|1\rangle )}{\sqrt{2}}%
\longrightarrow \mid \Phi ^{\pm }\rangle _{A1-A2}\frac{(|0\rangle +|1\rangle
)}{\sqrt{2}},
\end{equation}%
and if we send atoms $A1$ and $A2$ through $C$ in the state (\ref{PSI+Casc})
or (\ref{PSI-Casc}) we have%
\begin{equation}
\mid \Psi ^{\pm }\rangle _{A1-A2}\frac{(|0\rangle +|1\rangle )}{\sqrt{2}}%
\longrightarrow \mid \Psi ^{\pm }\rangle _{A1-A2}\frac{(|0\rangle -|1\rangle
)}{\sqrt{2}}.
\end{equation}%
Now if we send an atom $A5$ through $C$ in the state $|f_{5}\rangle $,
resonant with the cavity, with $|f_{5}\rangle $ and $|e_{5}\rangle $ being
the lower and upper level respectively, for $gt=\pi /2$, we have%
\begin{eqnarray}
|f_{5}\rangle (|0\rangle +|1\rangle ) &\longrightarrow &(|f_{5}\rangle
-i|e_{5}\rangle )|0\rangle , \\
|f_{5}\rangle (|0\rangle -|1\rangle ) &\longrightarrow &(|f_{5}\rangle
+i|e_{5}\rangle )|0\rangle ,  \nonumber
\end{eqnarray}%
Now we send atom $A5$ through a Ramsey cavity $R$, where the states are
rotated according to the rotation matrix (\ref{Rot2}), that is, we have%
\begin{eqnarray}
\frac{1}{\sqrt{2}}(i &\mid &e_{5}\rangle +\mid f_{5}\rangle )\rightarrow
i\mid e_{5}\rangle ,  \nonumber \\
\frac{1}{\sqrt{2}}(-i &\mid &e_{5}\rangle +\mid f_{5}\rangle )\rightarrow
\mid f_{5}\rangle ,
\end{eqnarray}%
Therefore, the detection of $\mid f_{5}\rangle $ corresponds to the
detection of \ $\mid \Phi ^{\pm }\rangle _{A1-A2}$ and $\ $of $\mid
e_{5}\rangle $ \ corresponds to the detection of \ $\mid \Psi ^{\pm }\rangle
_{A1-A2}$. Now we have to distinguish $(\mid \Psi ^{+}\rangle _{A1-A2},\mid
\Phi ^{+}\rangle _{A1-A2})$ from $(\mid \Psi ^{-}\rangle _{A1-A2},\mid \Phi
^{-}\rangle _{A1-A2})$. In order to do this we notice that, defining 
\begin{equation}
\Sigma _{x}=\sigma _{x}^{1}\sigma _{x}^{2},
\end{equation}%
where%
\begin{equation}
\sigma _{x}^{k}=\mid f_{k}\rangle \langle g_{k}\mid +\mid g_{k}\rangle
\langle f_{k}\mid ,
\end{equation}%
we have%
\begin{eqnarray}
\Sigma _{x} &\mid &\Psi ^{\pm }\rangle _{A1-A2}=\pm \mid \Psi ^{\pm }\rangle
_{A1-A2},  \nonumber \\
\Sigma _{x} &\mid &\Phi ^{\pm }\rangle _{A1-A2}=\pm \mid \Phi ^{\pm }\rangle
_{A1-A2}.  \label{AVSIGMAx}
\end{eqnarray}%
Therefore, we can distinguish between $(\mid \Psi ^{+}\rangle _{A1-A2},\mid
\Phi ^{+}\rangle _{A1-A2})$ and $(\mid \Psi ^{-}\rangle _{A1-A2},\mid \Phi
^{-}\rangle _{A1-A2})$ performing measurements of $\Sigma _{x}=\sigma
_{x}^{1}\sigma _{x}^{2}$. In order to do so we proceed as follows. We make
use of

\begin{equation}
K_{k}=\frac{1}{\sqrt{2}}\left[ 
\begin{array}{cc}
1 & -1 \\ 
1 & 1%
\end{array}%
\right] ,
\end{equation}%
or%
\begin{equation}
K_{k}=\frac{1}{\sqrt{2}}(\mid f_{k}\rangle \langle f_{k}\mid -\mid
f_{k}\rangle \langle g_{k}\mid +\mid g_{k}\rangle \langle f_{k}\mid +\mid
g_{k}\rangle \langle g_{k}\mid ),  \label{KkEPR}
\end{equation}%
to gradually unravel the Bell states. The eigenvectors of the operators $%
\sigma _{x}^{k}$ are%
\begin{equation}
|\psi _{x}^{k},\pm \rangle =\frac{1}{\sqrt{2}}(\mid f_{k}\rangle \pm \mid
g_{k}\rangle ),  \label{PSIxEPR}
\end{equation}%
and we can rewrite the Bell states as 
\begin{eqnarray}
&\mid &\Phi ^{\pm }\rangle _{A1-A2}=\frac{1}{2}[|\psi _{x}^{1},+\rangle
(\mid f_{2}\rangle \pm \mid g_{2}\rangle +|\psi _{x}^{1},-\rangle (\mid
f_{2}\rangle \mp \mid g_{2}\rangle )],  \nonumber \\
&\mid &\Psi ^{\pm }\rangle _{A1-A2}=\frac{1}{2}[|\psi _{x}^{1},+\rangle
(\mid g_{2}\rangle \pm \mid f_{2}\rangle )+|\psi _{x}^{1},-\rangle (\mid
g_{2}\rangle \mp \mid f_{2}\rangle ].  \label{EPRPSIx}
\end{eqnarray}%
Let us take for instance (\ref{PHI+Casc}) 
\begin{equation}
\mid \Phi ^{+}\rangle _{A1-A2}=\frac{1}{\sqrt{2}}(\mid f_{1}\rangle \mid
f_{2}\rangle +\mid g_{1}\rangle \mid g_{2}\rangle ).
\end{equation}%
Applying $K_{1}$ to this state we have%
\begin{equation}
K_{1}\mid \Phi ^{+}\rangle _{A1-A2}=\frac{1}{2}[|f_{1}\rangle (\mid
f_{2}\rangle -\mid g_{2}\rangle )+|g_{1}\rangle (\mid f_{2}\rangle +\mid
g_{2}\rangle )].  \label{EPRK1PSI12}
\end{equation}%
Now, we compare (\ref{EPRK1PSI12}) and (\ref{EPRPSIx}). We see that the
rotation by $K_{1}$ followed by the detection of $|g_{1}\rangle $
corresponds to the detection of the the state $|\psi _{x}^{1},+\rangle $
whose eigenvalue of $\sigma _{x}^{1}$ is $+1$. After we detect $%
|g_{1}\rangle $, we get%
\begin{equation}
\mid \psi \rangle _{A2}=\frac{1}{\sqrt{2}}(\mid f_{2}\rangle +\mid
g_{2}\rangle ),
\end{equation}%
that is, we have got 
\begin{equation}
\mid \psi \rangle _{A2}=|\psi _{x}^{2},+\rangle .  \label{EPRPSI2x}
\end{equation}%
If we apply (\ref{KkEPR}) for $k=2$ to the state (\ref{EPRPSI2x}) we get%
\begin{equation}
K_{2}\mid \psi \rangle _{A2}=|g_{2}\rangle .
\end{equation}%
We see that the rotation by $K_{2}$ followed by the detection of $%
|g_{2}\rangle $ corresponds to the detection of the the state $|\psi
_{x}^{2},+\rangle $ whose eigenvalue of $\sigma _{x}^{2}$ is $+1$. The same
applies to (\ref{PSI+Casc}).

Summarizing, we have two possible sequences of atomic state rotations
through $K_{k}$ and detections of $\mid f_{k}\rangle $ or $\mid g_{k}\rangle 
$ and the corresponding states $|\psi _{x}^{k},\pm \rangle $ where $k=1$ and 
$2$ which corresponds to the measurement of the eigenvalue $+1$ of the
operator $\Sigma _{x}$ given by (\ref{AVSIGMAx}) and the detection of (\ref%
{PHI+Casc}) or (\ref{PSI+Casc}) corresponds to 
\begin{eqnarray}
(K_{1}, &\mid &g_{1}\rangle )(K_{2},\mid g_{2}\rangle )\longleftrightarrow
|\psi _{x}^{1},+\rangle |\psi _{x}^{2},+\rangle ,  \nonumber \\
(K_{1}, &\mid &f_{1}\rangle )(K_{2},\mid f_{2}\rangle )\longleftrightarrow
|\psi _{x}^{1},-\rangle |\psi _{x}^{2},-\rangle .  \label{TestSIGMAx+}
\end{eqnarray}

Considering (\ref{PHI-Casc}) and (\ref{PSI-Casc}) we have 
\begin{eqnarray}
(K_{1}, &\mid &g_{1}\rangle )(K_{2},\mid f_{2}\rangle )\longleftrightarrow
|\psi _{x}^{1},+\rangle |\psi _{x}^{2},-\rangle ,  \nonumber \\
(K_{1}, &\mid &f_{1}\rangle )(K_{2},\mid g_{2}\rangle )\longleftrightarrow
|\psi _{x}^{1},-\rangle |\psi _{x}^{2},+\rangle ,  \label{TestSIGMAx-}
\end{eqnarray}%
which corresponds to the measurement of the eigenvalue $-1$ of the operator $%
\Sigma _{x}$ \ given by (\ref{AVSIGMAx}).

Let us now assume that Alice keeps with her the half of the Bell state (\ref%
{PHI+Casc}) consisting of atom $A2$ and Bob keeps with him the other half of
this Bell state, that is, atom $A1$. Then, they separate and let us assume
that they are far apart from each other. Later on, Alice decides to teleport
the state of an atom $A4$ prepared in an unknown state%
\begin{equation}
\mid \psi \rangle _{A4}=\zeta \mid f_{4}\rangle +\xi \mid g_{4}\rangle 
\label{PsiTelCasc}
\end{equation}%
to Bob. Now let us write \ the state formed by the direct product of the
Bell state and the unknown state $\mid \Phi ^{+}\rangle _{A1-A2}\mid \psi
\rangle _{A4}$, that is,%
\begin{eqnarray}
&\mid &\psi \rangle _{A1-A2-A4}=\frac{1}{\sqrt{2}}[\zeta (\mid f_{1}\rangle
\mid f_{2}\rangle \mid f_{4}\rangle +\mid g_{1}\rangle \mid g_{2}\rangle
\mid f_{4}\rangle )+  \nonumber \\
\xi ( &\mid &f_{1}\rangle \mid f_{2}\rangle \mid g_{4}\rangle +\mid
g_{1}\rangle \mid g_{2}\rangle \mid g_{4}\rangle )].  \label{TELPSIA1A2A4}
\end{eqnarray}%
First Alice prepares a cavity $C$ in the state (\ref{PsiC}). Taking into
account (\ref{UCasc}) with $\varphi =\pi ,$ after atoms $A2$ and $A4$ fly
through the cavity we have%
\begin{eqnarray}
&\mid &\psi \rangle _{A1-A2-A4-C}=\frac{1}{2}\{\zeta \lbrack \mid
f_{1}\rangle \mid f_{2}\rangle \mid f_{4}\rangle (|0\rangle +|1\rangle
)+\mid g_{1}\rangle \mid g_{2}\rangle \mid f_{4}\rangle (|0\rangle
-|1\rangle )]+  \nonumber \\
\xi \lbrack  &\mid &f_{1}\rangle \mid f_{2}\rangle \mid g_{4}\rangle
(|0\rangle -|1\rangle )+\mid g_{1}\rangle \mid g_{2}\rangle \mid
g_{4}\rangle (|0\rangle +|1\rangle )]\}.  \label{TPSIA1A2A4C}
\end{eqnarray}%
Now, making use of the Bell basis involving atom $A1$ and $A2$ we can
rewrite (\ref{TPSIA1A2A4C}) as%
\begin{eqnarray}
&\mid &\psi \rangle _{A1-A2-A4-C}=  \nonumber \\
\frac{1}{2\sqrt{2}}[ &\mid &\Phi ^{+}\rangle _{A2-A4}(\zeta \mid
f_{1}\rangle +\xi \mid g_{1}\rangle )(|0\rangle +|1\rangle )+  \nonumber \\
&\mid &\Phi ^{-}\rangle _{A2-A4}(\zeta \mid f_{1}\rangle -\xi \mid
g_{1}\rangle )(|0\rangle +|1\rangle )+  \nonumber \\
&\mid &\Psi ^{+}\rangle _{A2-A4}(\zeta \mid g_{1}\rangle +\xi \mid
f_{1}\rangle )(|0\rangle -|1\rangle )+  \nonumber \\
&\mid &\Psi ^{-}\rangle _{A2-A4}(\zeta \mid g_{1}\rangle -\xi \mid
f_{1}\rangle )(|0\rangle -|1\rangle )].  \label{PSIA1A2A4CEPR}
\end{eqnarray}%
Now Alice sends an atom $A5$ through $C$ in the state $|f_{5}\rangle $,
resonant with the cavity, with $|f_{5}\rangle $ and $|e_{5}\rangle $ being
the lower and upper level respectively. For $gt=\pi /2$, we have%
\begin{eqnarray}
|f_{5}\rangle (|0\rangle +|1\rangle ) &\longrightarrow &(|f_{5}\rangle
-i|e_{5}\rangle )|0\rangle , \\
|f_{5}\rangle (|0\rangle -|1\rangle ) &\longrightarrow &(|f_{5}\rangle
+i|e_{5}\rangle )|0\rangle ,  \nonumber
\end{eqnarray}%
and 
\begin{eqnarray}
&\mid &\psi \rangle _{A1-A2-A4-A5-C}=  \nonumber \\
\frac{1}{2\sqrt{2}}[ &\mid &\Phi ^{+}\rangle _{A2-A4}(\zeta \mid
f_{1}\rangle +\xi \mid g_{1}\rangle )(|f_{5}\rangle -i|e_{5}\rangle )+ 
\nonumber \\
&\mid &\Phi ^{-}\rangle _{A2-A4}(\zeta \mid f_{1}\rangle -\xi \mid
g_{1}\rangle )(|f_{5}\rangle -i|e_{5}\rangle )+  \nonumber \\
&\mid &\Psi ^{+}\rangle _{A2-A4}(\zeta \mid g_{1}\rangle +\xi \mid
f_{1}\rangle )(|f_{5}\rangle +i|e_{5}\rangle )+  \nonumber \\
&\mid &\Psi ^{-}\rangle _{A2-A4}(\zeta \mid g_{1}\rangle -\xi \mid
f_{1}\rangle )(|f_{5}\rangle +i|e_{5}\rangle )]|0\rangle .
\end{eqnarray}%
Now, Alice sends atom $A5$ through a Ramsey cavity $R6$, where the states
are rotated according to the rotation matrix (\ref{Rot2}), that is,%
\begin{eqnarray}
\frac{1}{\sqrt{2}}(i &\mid &e_{5}\rangle +\mid f_{5}\rangle )\rightarrow
i\mid e_{5}\rangle ,  \nonumber \\
\frac{1}{\sqrt{2}}(-i &\mid &e_{5}\rangle +\mid f_{5}\rangle )\rightarrow
\mid f_{5}\rangle .
\end{eqnarray}%
and we have%
\begin{eqnarray}
&\mid &\psi \rangle _{A1-A2-A4-A5-C}=  \nonumber \\
\frac{1}{2}[ &\mid &\Phi ^{+}\rangle _{A2-A4}(\zeta \mid f_{1}\rangle +\xi
\mid g_{1}\rangle )|f_{5}\rangle +  \nonumber \\
&\mid &\Phi ^{-}\rangle _{A2-A4}(\zeta \mid f_{1}\rangle -\xi \mid
g_{1}\rangle )|f_{5}\rangle +  \nonumber \\
i &\mid &\Psi ^{+}\rangle _{A2-A4}(\zeta \mid g_{1}\rangle +\xi \mid
f_{1}\rangle )|e_{5}\rangle +  \nonumber \\
i &\mid &\Psi ^{-}\rangle _{A2-A4}(\zeta \mid g_{1}\rangle -\xi \mid
f_{1}\rangle )|e_{5}\rangle ]|0\rangle .
\end{eqnarray}%
Then, if she detects the lower state $\mid f_{5}\rangle $ she gets%
\begin{equation}
\mid \psi \rangle _{A1-A2-A4}=\frac{1}{N}[\mid \Phi ^{+}\rangle
_{A2-A4}(\zeta \mid f_{1}\rangle +\xi \mid g_{1}\rangle )+\mid \Phi
^{-}\rangle _{A2-A4}(\zeta \mid f_{1}\rangle -\xi \mid g_{1}\rangle )],
\end{equation}%
where $N$ is a normalization constant. Now Alice follow the above
prescription in order to distinguish $\mid \Phi ^{+}\rangle _{A2-A4}$ from $%
\mid \Phi ^{-}\rangle _{A2-A4}.$ That is, she proceeds according to (\ref%
{TestSIGMAx+}) and (\ref{TestSIGMAx-}). If Alice gets $(K_{2},\mid
g_{2}\rangle )(K_{4},\mid g_{4}\rangle )$ or $(K_{2},\mid f_{2}\rangle
)(K_{4},\mid f_{4}\rangle )$ this corresponds to the detection \ of $\mid
\Phi ^{+}\rangle _{A2-A4}$ and Bob gets 
\begin{equation}
\mid \psi \rangle _{A1}=\zeta \mid f_{1}\rangle +\xi \mid g_{1}\rangle ,
\end{equation}%
and he has to do thing since he got the correct state (\ref{PsiTelCasc}). If
Alice gets $(K_{2},\mid f_{2}\rangle )(K_{4},\mid g_{4}\rangle )$ or $%
(K_{2},\mid g_{2}\rangle )(K_{4},\mid f_{4}\rangle )$ this corresponds to
the detection \ of $\mid \Phi ^{-}\rangle _{A2-A4}$ and Bob gets 
\begin{equation}
\mid \psi \rangle _{A1}=\zeta \mid f_{1}\rangle -\xi \mid g_{1}\rangle ,
\end{equation}%
and he has to apply a rotation in the Ramsey cavity $R7$ 
\begin{equation}
R=\left[ 
\begin{array}{cc}
1 & 0 \\ 
0 & -1%
\end{array}%
\right] ,
\end{equation}%
to get (\ref{PsiTelCasc}). If Alice detects the upper state $\mid
e_{5}\rangle $ she gets%
\begin{equation}
\mid \psi \rangle _{A1-A2-A4}=\frac{1}{N}[\mid \Psi ^{+}\rangle
_{A2-A4}(\zeta \mid g_{1}\rangle +\xi \mid f_{1}\rangle )+\mid \Psi
^{-}\rangle _{A2-A4}(-\zeta \mid g_{1}\rangle +\xi \mid f_{1}\rangle )],
\end{equation}%
Now Alice follow the above prescription in order to distinguish $\mid \Psi
^{+}\rangle _{A2-A4}$ from $\mid \Psi ^{-}\rangle _{A2-A4}.$ That is, she
proceeds according to (\ref{TestSIGMAx+}) and (\ref{TestSIGMAx-}). If Alice
gets $(K_{2},\mid g_{2}\rangle )(K_{4},\mid g_{4}\rangle )$ or $(K_{2},\mid
f_{2}\rangle )(K_{4},\mid f_{4}\rangle )$ this corresponds to the detection
\ of $\mid \Psi ^{+}\rangle _{A2-A4}$ and Bob gets 
\begin{equation}
\mid \psi \rangle _{A1}=\zeta \mid g_{1}\rangle +\xi \mid f_{1}\rangle ,
\end{equation}%
and he has to apply a rotation in the Ramsey cavity $R7$ 
\[
R=\left[ 
\begin{array}{cc}
0 & 1 \\ 
1 & 0%
\end{array}%
\right] ,
\]%
to get (\ref{PsiTelCasc}). If Alice gets $(K_{2},\mid f_{2}\rangle
)(K_{4},\mid g_{4}\rangle )$ or $(K_{2},\mid g_{2}\rangle )(K_{4},\mid
f_{4}\rangle )$ this corresponds to the detection \ of $\mid \Psi
^{-}\rangle _{A2-A4}$ and Bob gets 
\begin{equation}
\mid \psi \rangle _{A1}=-\zeta \mid g_{1}\rangle +\xi \mid f_{1}\rangle ,
\end{equation}%
and he has to apply a rotation in the Ramsey cavity $R7$ 
\begin{equation}
R=\left[ 
\begin{array}{cc}
0 & -1 \\ 
1 & 0%
\end{array}%
\right] ,
\end{equation}%
to get (\ref{PsiTelCasc}).

Notice that the original state (\ref{PsiTelCasc}) \ is destroyed in the end
of the teleportation process (it evolves to \ $\mid f_{4}\rangle $ or $\mid
g_{4}\rangle $) in accordance with the no-cloning theorem \cite{Nielsen,
Nocloning}. In Fig. 2 we present the scheme of the teleportation process we
have discussed above.

\subsection{\protect\bigskip ATOMS IN A LAMBDA CONFIGURATION\protect\bigskip}

Consider a three-level lambda atom (see Fig. 3) interacting with an
electromagnetic field inside a cavity $C$. The states of \ the atom $%
|a\rangle ,$ $|b\rangle $ and $|c\rangle ,$ are so that the $|a\rangle
\rightleftharpoons |c\rangle $ and $|a\rangle \rightleftharpoons |b\rangle $
transitions are in the far off resonance interaction limit. The time
evolution operator for the atom-field interaction  is given by \cite{Knight}
(see Appendix) 
\begin{equation}
U(\tau )=-e^{i\varphi a^{\dagger }a}|a\rangle \langle a|+\frac{1}{2}%
(e^{i\varphi a^{\dagger }a}+1)|b\rangle \langle b|+\frac{1}{2}(e^{i\varphi
a^{\dagger }a}-1)|b\rangle \langle c|\ +\frac{1}{2}(e^{i\varphi a^{\dagger
}a}-1)|c\rangle \langle b|+\frac{1}{2}(e^{i\varphi a^{\dagger
}a}+1)|c\rangle \langle c|,  \label{U1lambda}
\end{equation}%
where $a$ $(a^{\dagger })$ is the annihilation (creation) operator for the
field in cavity $C$, $\varphi =2g^{2}\tau /$ $\Delta $, \ $g$ is the
coupling constant, $\Delta =\omega _{a}-\omega _{b}-\omega =\omega
_{a}-\omega _{c}-\omega $ is the detuning where \ $\omega _{a}$, $\omega _{b}
$ and $\omega _{c}$\ are the frequency of the upper level and \ of the two
degenerate lower levels respectively and $\omega $ is the cavity field
frequency and $\tau $ is the atom-field interaction time. For $\varphi =\pi $%
, we get 
\begin{equation}
U(\tau )=-\exp \left( i\pi a^{\dagger }a\right) |a\rangle \langle a|+\Pi
_{+}|b\rangle \langle b|+\Pi _{-}|b\rangle \langle c|\ +\Pi _{-}|c\rangle
\langle b|+\Pi _{+}|c\rangle \langle c|,  \label{UlambdaPi}
\end{equation}%
where 
\begin{eqnarray}
\Pi _{+} &=&\frac{1}{2}(e^{i\pi a^{\dagger }a}+1),  \nonumber \\
\Pi _{-} &=&\frac{1}{2}(e^{i\pi a^{\dagger }a}-1).  \label{pi+-}
\end{eqnarray}

Let us first show how we can get Bell states making use of three-level
lambda atoms interacting with a cavity field prepared in state (\ref{PsiC}).
Consider an atom $A1$ in the state $|\psi \rangle _{A1}=|b_{1}\rangle $ and
a cavity $C$ prepared in the state (\ref{PsiC}). We now let atom $A1$ to fly
through the cavity $C$. Taking into account (\ref{UlambdaPi}) the state \ of
the system $A1-C$ evolves to 
\begin{equation}
|\psi \rangle _{A1-C}=\frac{1}{\sqrt{2}}(|b_{1}\rangle |0\rangle
-|c_{1}\rangle |1\rangle ).
\end{equation}%
Consider now another three-level lambda atom $A2$ prepared initially in the
state $|b_{2}\rangle $, which is going to pass through the cavity. After
this second atom has passed through the cavity, the system evolves to 
\begin{equation}
|\psi \rangle _{A1-A2-C}=\frac{1}{\sqrt{2}}(|b_{1}\rangle |b_{2}\rangle
|0\rangle +|c_{1}\rangle |c_{2}\rangle |1\rangle ).  \label{LBPSIA1A2C}
\end{equation}%
In order to disentangle the atomic states of the cavity field state we now
send a two-level atom $A3,$ resonant with the cavity, with $|f_{3}\rangle $
and $|e_{3}\rangle $ being the lower and upper levels respectively, through $%
C$. If $A3$ is sent in the lower state $|f_{3}\rangle $, under the
Jaynes-Cummings dynamics (see\ (\ref{UJC}), for $\Delta =0$ and $gt=\pi /2$)
we know that the state $|f_{3}\rangle |0\rangle $ does not evolve, however,
the state $|f_{3}\rangle |1\rangle $ evolves to $-i|e_{3}\rangle |0\rangle $%
. Then we get%
\begin{equation}
|\psi (\tau )\rangle _{A1-A2-A3-C}=\frac{1}{\sqrt{2}}(|b_{1}\rangle
|b_{2}\rangle |f_{3}\rangle -i|c_{1}\rangle |c_{2}\rangle |e_{3}\rangle
)|0\rangle .
\end{equation}%
Now we let atom $A3$ to enter a Ramsey cavity $R1$ where the atomic states
are rotated according to the rotation matrix (\ref{Rot2}), that is,%
\begin{eqnarray}
&\mid &e_{3}\rangle \rightarrow \frac{1}{\sqrt{2}}(\mid e_{3}\rangle +i\mid
f_{3}\rangle ),  \nonumber \\
&\mid &f_{3}\rangle \rightarrow \frac{1}{\sqrt{2}}(i\mid e_{3}\rangle +\mid
f_{3}\rangle ),
\end{eqnarray}%
and we get%
\begin{equation}
|\psi \rangle _{A1-A2-A3}=\frac{1}{2}[|b_{1}\rangle |b_{2}\rangle
|f_{3}\rangle (i\mid e_{3}\rangle +\mid f_{3}\rangle )-i|c_{1}\rangle
|c_{2}\rangle (\mid e_{3}\rangle +i\mid f_{3}\rangle )].
\end{equation}%
and if we detect atom $A3$ in state $|f_{3}\rangle $ finally we get the the
Bell state 
\begin{equation}
\mid \Phi ^{+}\rangle _{A1-A2}=\frac{1}{\sqrt{2}}(|b_{1}\rangle
|b_{2}\rangle +|c_{1}\rangle |c_{2}\rangle ),  \label{PHI+Lamb}
\end{equation}%
which is an entangled state of atoms $A1$ and $A2$. If we detect atom $A3$
in state $|e_{3}\rangle $ we get 
\begin{equation}
\mid \Phi ^{-}\rangle _{A1-A2}=\frac{1}{\sqrt{2}}(\mid b_{1}\rangle \mid
b_{2}\rangle -\mid c_{1}\rangle \mid c_{2}\rangle ),  \label{PHI-Lamb}
\end{equation}%
Now, if we apply an extra rotation on the states of atom $A2$ in (\ref%
{PHI+Lamb}) in a Ramsey cavity $R,$ according to the rotation matrix%
\begin{equation}
R=\mid b_{2}\rangle \langle c_{2}|+\mid c_{2}\rangle \langle b_{2}|,
\label{R}
\end{equation}%
we get%
\begin{equation}
\mid \Psi ^{+}\rangle _{A1-A2}=\frac{1}{\sqrt{2}}(\mid b_{1}\rangle \mid
c_{2}\rangle +\mid c_{1}\rangle \mid b_{2}\rangle ),  \label{PSI+Lamb}
\end{equation}%
and applying (\ref{R}) to (\ref{PHI-Lamb}) we get%
\begin{equation}
\mid \Psi ^{-}\rangle _{A1-A2}=\frac{1}{\sqrt{2}}(\mid b_{1}\rangle \mid
c_{2}\rangle -\mid c_{1}\rangle \mid b_{2}\rangle ).  \label{PSI-Lamb}
\end{equation}

Now, let us assume that Alice keeps with her the half of the Bell state (\ref%
{PHI+Lamb}) consisting of atom $A2$ and Bob keeps with him the other half of
this Bell state, that is, atom $A1$. Then they separate and let us assume
that they are far apart from each other. Later on, Alice decides to teleport
the state of an atom $A4$ prepared in an unknown state%
\begin{equation}
\mid \psi \rangle _{A4}=\zeta \mid b_{4}\rangle +\xi \mid c_{4}\rangle
\label{PsiTelLamb}
\end{equation}%
to Bob. First Alice prepares a cavity $C$ in the state (\ref{PsiC})$.$ Let
us write \ the state formed by the direct product of the Bell state and the
unknown state $\mid \Phi ^{+}\rangle _{A1-A2}\mid \psi \rangle _{A4}$, that
is,%
\begin{eqnarray}
&\mid &\psi \rangle _{A1-A2-A4}=\frac{1}{\sqrt{2}}[\zeta (\mid b_{1}\rangle
\mid b_{2}\rangle \mid b_{4}\rangle +\mid c_{1}\rangle \mid c_{2}\rangle
\mid b_{4}\rangle )+  \nonumber \\
\xi ( &\mid &b_{1}\rangle \mid b_{2}\rangle \mid c_{4}\rangle +\mid
c_{1}\rangle \mid c_{2}\rangle \mid c_{4}\rangle )].
\end{eqnarray}%
Taking into account (\ref{UlambdaPi}), after Alice let atoms $A2$ and $A4$
to fly through $C$ prepared in the state (\ref{PsiC}), she gets%
\begin{eqnarray}
&\mid &\psi \rangle _{A1-A2-A4-C}=  \nonumber \\
\frac{1}{2\sqrt{2}}[ &\mid &\Phi ^{+}\rangle _{A2-A4}(\zeta \mid
b_{1}\rangle +\xi \mid c_{1}\rangle )(|0\rangle +|1\rangle )+  \nonumber \\
&\mid &\Phi ^{-}\rangle _{A2-A4}(\zeta \mid b_{1}\rangle -\xi \mid
c_{1}\rangle )(|0\rangle -|1\rangle )+  \nonumber \\
&\mid &\Psi ^{+}\rangle _{A2-A4}(\zeta \mid c_{1}\rangle +\xi \mid
b_{1}\rangle )(|0\rangle +|1\rangle )+  \nonumber \\
&\mid &\Psi ^{-}\rangle _{A2-A4}(-\zeta \mid c_{1}\rangle +\xi \mid
b_{1}\rangle )(|0\rangle -|1\rangle )].
\end{eqnarray}%
In order to disentangle the atomic states of the cavity field state Alice
sends a two-level atom $A5,$ resonant with the cavity, with $|f_{5}\rangle $
and $|e_{5}\rangle $ being the lower and upper levels respectively, through $%
C$. If $A5$ is sent in the lower state $|f_{5}\rangle $, under the
Jaynes-Cummings dynamics (see\ (\ref{UJC}), for $\Delta =0$ and $gt=\pi /2$)
we know that the state $|f_{5}\rangle |0\rangle $ does not evolve, however,
the state $|f_{5}\rangle |1\rangle $ evolves to $-i|e_{5}\rangle |0\rangle $
and therefore she gets%
\begin{eqnarray}
&\mid &\psi \rangle _{A1-A2-A4-A5-C}=  \nonumber \\
\frac{1}{2\sqrt{2}}[ &\mid &\Phi ^{+}\rangle _{A2-A4}(\zeta \mid
b_{1}\rangle +\xi \mid c_{1}\rangle )(|f_{5}\rangle -i|e_{5}\rangle )+ 
\nonumber \\
&\mid &\Phi ^{-}\rangle _{A2-A4}(\zeta \mid b_{1}\rangle -\xi \mid
c_{1}\rangle )(|f_{5}\rangle +i|e_{5}\rangle )+  \nonumber \\
&\mid &\Psi ^{+}\rangle _{A2-A4}(\zeta \mid c_{1}\rangle +\xi \mid
b_{1}\rangle )(|f_{5}\rangle -i|e_{5}\rangle )+  \nonumber \\
&\mid &\Psi ^{-}\rangle _{A2-A4}(-\zeta \mid c_{1}\rangle +\xi \mid
b_{1}\rangle )(|f_{5}\rangle +i|e_{5}\rangle )]|0\rangle .
\end{eqnarray}%
Now, if atom $A5$ enters a Ramsey cavity $R2$ where the atomic states are
rotated according to the rotation matrix (\ref{Rot2}), we have%
\begin{eqnarray}
\frac{1}{\sqrt{2}}(i &\mid &e_{5}\rangle +\mid f_{5}\rangle )\rightarrow
i\mid e_{5}\rangle ,  \nonumber \\
\frac{1}{\sqrt{2}}(-i &\mid &e_{5}\rangle +\mid f_{5}\rangle )\rightarrow
\mid f_{5}\rangle .
\end{eqnarray}%
and we get%
\begin{eqnarray}
&\mid &\psi \rangle _{A1-A2-A4-A5}=  \nonumber \\
\frac{1}{2}[ &\mid &\Phi ^{+}\rangle _{A2-A4}(\zeta \mid b_{1}\rangle +\xi
\mid c_{1}\rangle )|f_{5}\rangle +  \nonumber \\
i &\mid &\Phi ^{-}\rangle _{A2-A4}(\zeta \mid b_{1}\rangle -\xi \mid
c_{1}\rangle )|e_{5}\rangle +  \nonumber \\
&\mid &\Psi ^{+}\rangle _{A2-A4}(\zeta \mid c_{1}\rangle +\xi \mid
b_{1}\rangle )|f_{5}\rangle +  \nonumber \\
i &\mid &\Psi ^{-}\rangle _{A2-A4}(-\zeta \mid c_{1}\rangle +\xi \mid
b_{1}\rangle )|e_{5}\rangle ].
\end{eqnarray}%
If Alice detects $\mid f_{5}\rangle $ she gets%
\begin{equation}
\mid \psi \rangle _{A1-A2-A4}=\frac{1}{N}[\mid \Phi ^{+}\rangle
_{A2-A4}(\zeta \mid b_{1}\rangle +\xi \mid c_{1}\rangle )+\mid \Psi
^{+}\rangle _{A2-A4}(\zeta \mid c_{1}\rangle +\xi \mid b_{1}\rangle )].
\end{equation}%
where $N$ is a normalization constant. If Alice detects $(\mid b_{2}\rangle
\mid b_{4}\rangle )$ or $(\mid c_{2}\rangle \mid c_{4}\rangle )$ Bob gets 
\begin{equation}
\mid \psi \rangle _{A1}=\zeta \mid b_{1}\rangle +\xi \mid c_{1}\rangle ,
\end{equation}%
and he has to do nothing since this is the correct state (\ref{PsiTelLamb}).
If she detects $(\mid b_{2}\rangle \mid c_{4}\rangle )$ or $(\mid
c_{2}\rangle \mid b_{4}\rangle )$ Bob gets 
\begin{equation}
\mid \psi \rangle _{A1}=\zeta \mid c_{1}\rangle +\xi \mid b_{1}\rangle ,
\end{equation}%
and he has to apply a rotation in the Ramsey cavity $R3$ 
\begin{equation}
R_{3}=\left[ 
\begin{array}{cc}
0 & 1 \\ 
1 & 0%
\end{array}%
\right] ,
\end{equation}%
to get (\ref{PsiTelLamb}). If Alice detects $\mid e_{5}\rangle $ she gets%
\begin{equation}
\mid \psi \rangle _{A1-A2-A4}=\frac{i}{N}[\mid \Phi ^{-}\rangle
_{A2-A4}(\zeta \mid b_{1}\rangle -\xi \mid c_{1}\rangle )+\mid \Psi
^{-}\rangle _{A2-A4}(-\zeta \mid c_{1}\rangle +\xi \mid b_{1}\rangle )].
\end{equation}%
If Alice detects $(\mid b_{2}\rangle \mid b_{4}\rangle )$ or $(\mid
c_{2}\rangle \mid c_{4}\rangle )$ Bob gets 
\begin{equation}
\mid \psi \rangle _{A1}=\zeta \mid c_{1}\rangle -\xi \mid b_{1}\rangle ,
\end{equation}%
and he has to apply a rotation in the Ramsey cavity $R3$ 
\begin{equation}
R=\left[ 
\begin{array}{cc}
0 & 1 \\ 
-1 & 0%
\end{array}%
\right] ,
\end{equation}%
to get (\ref{PsiTelLamb}). If she detects $(\mid b_{2}\rangle \mid
c_{4}\rangle )$ or $(\mid c_{2}\rangle \mid b_{4}\rangle )$ Bob gets 
\begin{equation}
\mid \psi \rangle _{A1}=-\zeta \mid c_{1}\rangle +\xi \mid b_{1}\rangle ,
\end{equation}%
and he has to apply a rotation in the Ramsey cavity $R3$ 
\begin{equation}
R=\left[ 
\begin{array}{cc}
0 & -1 \\ 
1 & 0%
\end{array}%
\right] ,
\end{equation}%
to get (\ref{PsiTelLamb}). In Fig. 4 we present the scheme of the
teleportation process we have discussed above.

\section{\protect\bigskip CONCLUSION}

Concluding, we have presented two schemes of realization of atomic state
teleportation making use of cavity QED.\ In the schemes presented here we
use atoms interacting with a superconducting cavity prepared in a state $%
(|0\rangle +|1\rangle )/\sqrt{2}$ which is a state relatively easy to be
prepared and handled and for which decoherence is not so drastic. In the
first scheme we make use of atoms in a cascade configuration and in the
second scheme we make use of atoms in a lambda configuration. The advantage
of using a cascade atomic configuration is that the atomic state detection
process is simpler than in the lambda configuration where we have states
which are degenerated. On the other hand, for the cascade configuration we
have to perform more atomic state rotations using Ramsey cavities than in
the case of the lambda configuration. Nice alternative schemes also making
use of atoms interacting with electromagnetic cavities have also been
proposed in Refs. \cite{GuerraTelCascatCS, GuerraTelLambatCS, DavZag}.

\appendix

\section{Time evolution operator}

\bigskip

\subsection{Two-level atoms\protect\bigskip}

Let us consider a two-level atom interacting with a cavity field, where $%
|e\rangle $ and $|f\rangle $ are the upper and lower states respectively,
with $\omega _{e}$ and $\omega _{f}$ being the two atomic frequencies
associated to these two states and $\omega $ the cavity field frequency (see
Fig. 1). The Jaynes-Cummings Hamiltonian, under the rotating-wave
approximation, is given by 
\begin{equation}
H=\hbar a^{\dag }a+\hbar \omega _{e}|e\rangle \langle e|++\hbar \omega
_{f}|f\rangle \langle f|+\hbar g[a|e\rangle \langle f|+a^{\dag }|f\rangle
\langle e|),  \label{JCH}
\end{equation}%
where $a^{\dag }$ and $a$ are the creation and annihilation operators
respectively for the cavity field, $g$ is the coupling constant. We write%
\begin{equation}
H=H_{0}+H_{I},
\end{equation}%
where we have settled%
\begin{eqnarray}
H_{0} &=&\hbar a^{\dag }a+\hbar \omega _{e}|e\rangle \langle e|++\hbar
\omega _{f}|f\rangle \langle f|,  \nonumber \\
H_{I} &=&\hbar g[a|e\rangle \langle f|+a^{\dag }|f\rangle \langle e|).
\end{eqnarray}%
Lets define the interaction picture%
\begin{equation}
|\psi _{I}\rangle =e^{i\frac{H_{0}}{\hbar }t}|\psi _{S}\rangle ,
\end{equation}%
where%
\begin{equation}
i\hbar \frac{d}{dt}|\psi _{S}\rangle =H|\psi _{S}\rangle .
\end{equation}%
Then, we get%
\begin{equation}
i\hbar \frac{d}{dt}|\psi _{I}\rangle =V_{I}|\psi _{I}\rangle ,
\end{equation}%
where%
\begin{equation}
V_{I}=e^{i\frac{H_{0}}{\hbar }t}H_{I}e^{-i\frac{H_{0}}{\hbar }t}=\hbar \left[
\begin{array}{cc}
0 & ge^{i\Delta t}a \\ 
ge^{-i\Delta t}a^{\dagger } & 0%
\end{array}%
\right]
\end{equation}%
and%
\begin{equation}
\Delta =(\omega _{e}-\omega _{f})-\omega .
\end{equation}%
Considering%
\begin{equation}
|\psi _{I}(t)\rangle =U_{I}(t)|\psi _{I}(0)\rangle =U_{I}(t)|\psi
_{S}(0)\rangle ,
\end{equation}%
we have to solve the Schr\"{o}dinger's equation for the time evolution
operator%
\begin{equation}
i\hbar \frac{dU_{I}}{dt}=V_{I}U_{I},  \label{AP2}
\end{equation}%
where%
\begin{equation}
U_{I}(t)=\left[ 
\begin{array}{cc}
u_{ee}(t) & u_{ef}(t) \\ 
u_{fe}(t) & u_{ff}(t)%
\end{array}%
\right]
\end{equation}%
and%
\begin{equation}
U_{I}(0)=\left[ 
\begin{array}{cc}
1 & 0 \\ 
0 & 1%
\end{array}%
\right] .
\end{equation}%
That is,%
\begin{eqnarray}
i\frac{d}{dt}u_{ee}(t) &=&ge^{i\Delta t}au_{ef}(t),  \nonumber \\
i\frac{d}{dt}u_{ef}(t) &=&ge^{i\Delta t}au_{ff}(t),  \nonumber \\
i\frac{d}{dt}u_{fe}(t) &=&ge^{-i\Delta t}a^{\dagger }u_{ee}(t),  \nonumber \\
i\frac{d}{dt}u_{ff}(t) &=&ge^{-i\Delta t}a^{\dagger }u_{ef}(t),  \label{EDI4}
\end{eqnarray}%
which can be solved easily using, for instance, Laplace transformation, and
we get%
\begin{equation}
U_{I}(t)=\left[ 
\begin{array}{cc}
e^{i\frac{\Delta }{2}t}(\cos \mu t-i\frac{\Delta }{2\mu }\sin \mu t) & 
-ige^{i\frac{\Delta }{2}t}\frac{1}{\mu }(\sin \mu t)a \\ 
-iga^{\dag }e^{-i\frac{\Delta }{2}t}\frac{1}{\mu }(\sin \mu t) & e^{-i\frac{%
\Delta }{2}t}(\cos \nu t+i\frac{\Delta }{2\nu }\sin \nu t)%
\end{array}%
\right] ,  \label{UJC}
\end{equation}%
where we have defined%
\begin{eqnarray}
\mu &=&\sqrt{g^{2}aa^{\dag }+\frac{\Delta ^{2}}{4}},  \nonumber \\
\nu &=&\sqrt{g^{2}a^{\dag }a+\frac{\Delta ^{2}}{4}}.  \label{AP4}
\end{eqnarray}

In the large detuning limit ($\Delta \gg g$) we have%
\begin{eqnarray}
\mu &=&\sqrt{g^{2}aa^{\dag }+\frac{\Delta ^{2}}{4}}\cong \frac{\Delta }{2}+%
\frac{g^{2}aa^{\dag }}{\Delta },  \nonumber \\
\nu &=&\sqrt{g^{2}a^{\dag }a+\frac{\Delta ^{2}}{4}}.\cong \frac{\Delta }{2}+%
\frac{g^{2}a^{\dag }a}{\Delta }.
\end{eqnarray}%
and we get easily 
\begin{equation}
U_{d}(t)=e^{-i\varphi (a^{\dagger }a+1)}\mid e\rangle \langle e\mid
+e^{i\varphi a^{\dagger }a}\mid f\rangle \langle f\mid ,  \label{Ud}
\end{equation}%
where $\varphi =$ $g^{2}t/\Delta .$

In the case we have a resonant interaction of an atom with cavity field ($%
\Delta =0$ in (\ref{UJC})), if the field is a very intense field we can
treat it classically. That is, in the time evolution operator (\ref{UJC}) \
we set $\Delta =0,$ and we substitute the creation and annihilation field
operators according to $a\rightarrow \eta e^{i\theta }$ and $a^{\dag
}\rightarrow \eta e^{-i\theta }$ where $\eta $ and $\theta $ are c-numbers.
Then, we have a semiclassical approach of the atom-field interaction in
which the field is treated classically and the atoms according to quantum
mechanics. In this case (\ref{UJC}) becomes%
\begin{equation}
U_{I,SC}(t)=\left[ 
\begin{array}{cc}
\cos (g\eta t) & -ie^{i\theta }\sin (g\eta t) \\ 
-ie^{-i\theta }\sin (g\eta t) & \cos (g\eta t)%
\end{array}%
\right] .  \label{URam}
\end{equation}%
Now we take $\theta =\pi /2$. If we choose $g\eta t=\pi /4$ we have the
rotation matrix%
\begin{equation}
R_{\frac{\pi }{2},\frac{\pi }{4}}=\frac{1}{\sqrt{2}}\left[ 
\begin{array}{cc}
1 & 1 \\ 
-1 & 1%
\end{array}%
\right] ,
\end{equation}%
and for $g\eta t=-\pi /4$ we have%
\begin{equation}
R_{\frac{\pi }{2},-\frac{\pi }{4}}=\frac{1}{\sqrt{2}}\left[ 
\begin{array}{cc}
1 & -1 \\ 
1 & 1%
\end{array}%
\right] .
\end{equation}%
For $g\eta t=\pi /2$ we get 
\begin{equation}
R_{\frac{\pi }{2},\frac{\pi }{2}}=\left[ 
\begin{array}{cc}
0 & 1 \\ 
-1 & 0%
\end{array}%
\right] ,
\end{equation}%
and for $g\eta t=-\pi /2$ 
\begin{equation}
R_{\frac{\pi }{2},-\frac{\pi }{2}}=\left[ 
\begin{array}{cc}
0 & -1 \\ 
1 & 0%
\end{array}%
\right] .
\end{equation}%
Now if we take $\theta =\pi $ and $g\eta t=\pi /4$ we have the rotation
matrix%
\begin{equation}
R_{\pi ,\frac{\pi }{4}}=\frac{1}{\sqrt{2}}\left[ 
\begin{array}{cc}
1 & i \\ 
i & 1%
\end{array}%
\right] ,
\end{equation}%
Choosing the proper values of $\theta $, $g$, $\eta $ and $t$ we can get the
rotation matrix we need to perform the rotation of the atomic states we
desire in a Ramsey cavity.

Just as a remark, consider (\ref{Ud}) and assume we have an intense field,
that is, we can use a semiclassical approach. In this case we set $%
a\rightarrow \eta e^{i\theta }$ and $a^{\dag }\rightarrow \eta e^{-i\theta }$
where $\eta $ and $\theta $ are c-numbers and $a^{\dag }a\rightarrow \eta
^{2}$. Defining $\varphi a^{\dagger }a\rightarrow \varphi \eta ^{2}=\beta /2$%
, (\ref{Ud}) reads%
\begin{equation}
U_{d,SC}=\left[ 
\begin{array}{cc}
e^{-i\beta /2} & 0 \\ 
0 & e^{i\beta /2}%
\end{array}%
\right] ,  \label{Ud,SC}
\end{equation}%
and for $\beta =\pi $ we have%
\begin{equation}
U_{\pi }=i\left[ 
\begin{array}{cc}
-1 & 0 \\ 
0 & 1%
\end{array}%
\right] .
\end{equation}%
Any arbitrary $2\times 2$ unitary matrix \ $M$\ may be decomposed as \cite%
{Nielsen} 
\begin{equation}
M=e^{i\alpha }\left[ 
\begin{array}{cc}
e^{-i\beta /2} & 0 \\ 
0 & e^{i\beta /2}%
\end{array}%
\right] \left[ 
\begin{array}{cc}
\cos \frac{\gamma }{2} & -\sin \frac{\gamma }{2} \\ 
\sin \frac{\gamma }{2} & \cos \frac{\gamma }{2}%
\end{array}%
\right] \left[ 
\begin{array}{cc}
e^{-i\delta /2} & 0 \\ 
0 & e^{i\delta /2}%
\end{array}%
\right] ,
\end{equation}%
where $\alpha ,\beta ,\gamma $ and $\delta $ are real parameters. Therefore,
we can use (\ref{URam}) and (\ref{Ud,SC}) to get a rotation matrix we need.

\subsection{Three-level lambda atoms}

We start with the Hamiltonian of a degenerate three-level lambda atom (see
Fig. 3) interacting with a field cavity mode

\begin{equation}
H=\hbar \omega a^{\dagger }a+\hbar \omega _{a}|a\rangle \langle a|\ +\hbar
\omega _{b}|b\rangle \langle b|+\hbar \omega _{c}|c\rangle \langle c|\
+\hbar a(g_{1}|a\rangle \langle b|+g_{2}|a\rangle \langle c|)+\hbar
a^{\dagger }(g_{1}^{\ast }|b\rangle \langle a|+g_{2}^{\ast }|c\rangle
\langle a|),
\end{equation}%
where $|a\rangle $, $|b\rangle $ and $|c\rangle $ are the upper and the two
degenerated lower atomic levels respectively, $a$ ($a^{\dagger }$) is the
annihilation (creation) field operator and $g_{1}$ and $g_{2}$ are the
coupling constants corresponding to the transitions $|a\rangle
\rightleftharpoons |c\rangle $ and $|a\rangle \rightleftharpoons |b\rangle $%
, respectively. In the interaction picture, 
\begin{equation}
V=\hbar \left[ e^{i\Delta _{1}t}g_{1}a\mid a\rangle \langle b\mid
+e^{-i\Delta _{1}t}g_{1}^{\ast }a^{\dagger }\mid b\rangle \langle a\mid %
\right] +\hbar \left[ e^{i\Delta _{2}t}g_{2}a\mid a\rangle \langle c\mid
+e^{-i\Delta _{2}t}g_{2}^{\ast }a^{\dagger }\mid c\rangle \langle a\mid %
\right] ,
\end{equation}%
where 
\begin{eqnarray}
\Delta _{1} &=&\omega _{a}-\omega _{b}-\omega \ \   \nonumber \\
\ \Delta _{2} &=&\omega _{a}-\omega _{c}-\omega .
\end{eqnarray}%
The time evolution operator 
\begin{equation}
U(t)=\left[ 
\begin{array}{lll}
u_{aa} & u_{ab} & u_{ac} \\ 
u_{ba} & u_{bb} & u_{bc} \\ 
u_{aa} & u_{cb} & u_{cc}%
\end{array}%
\right] ,
\end{equation}%
whose initial condition is given by 
\begin{equation}
U(0)=\left[ 
\begin{array}{lll}
1 & 0 & 0 \\ 
0 & 1 & 0 \\ 
0 & 0 & 1%
\end{array}%
\right] ,
\end{equation}%
should satisfy the Schr\"{o}dinger equation of motion 
\begin{equation}
i\hbar \frac{dU}{dt}=VU=\left[ 
\begin{array}{lll}
\zeta _{1}u_{ba}+\zeta _{2}u_{ca} & \zeta _{1}u_{bb}+\zeta _{2}u_{cb} & 
\zeta _{1}u_{bc}+\zeta _{2}u_{cc} \\ 
\zeta _{1}^{\dagger }u_{aa} & \zeta _{1}^{\dagger }u_{ab} & \zeta
_{1}^{\dagger }u_{ac} \\ 
\zeta _{2}^{\dagger }u_{aa} & \zeta _{2}^{\dagger }u_{ab} & \zeta
_{2}^{\dagger }u_{ac}%
\end{array}%
\right] ,
\end{equation}%
where 
\begin{eqnarray}
\zeta _{1} &=&\hbar e^{i\Delta _{1}t}g_{1}a,  \nonumber \\
\zeta _{2} &=&\hbar e^{i\Delta _{2}t}g_{2}a.
\end{eqnarray}%
Observe that this equation may be grouped in three sets of couple
differential equations. One for $u_{aa}$, $u_{ba}$ and $u_{ca}$, i.e., 
\begin{eqnarray}
i\hbar \frac{du_{aa}}{dt} &=&\zeta _{1}u_{ba}+\zeta _{2}u_{ca},\hspace{8pt} 
\nonumber \\
i\hbar \frac{du_{ba}}{dt} &=&\zeta _{1}^{\dagger }u_{aa},  \nonumber \\
i\hbar \frac{du_{ca}}{dt} &=&\zeta _{2}^{\dagger }u_{aa},
\end{eqnarray}%
another involving only $u_{ab}$, $u_{bb}$ and $u_{cb}$, 
\begin{eqnarray}
i\hbar \frac{du_{ab}}{dt} &=&\zeta _{1}u_{bb}+\zeta _{2}u_{cb},\hspace{8pt} 
\nonumber \\
i\hbar \frac{du_{bb}}{dt} &=&\zeta _{1}^{\dagger }u_{ab},  \nonumber \\
i\hbar \frac{du_{cb}}{dt} &=&\zeta _{2}^{\dagger }u_{ab},
\end{eqnarray}%
and, finally, one involving $u_{ac}$, $u_{bc}$ and $u_{cc}$, 
\begin{eqnarray}
i\hbar \frac{du_{ac}}{dt} &=&\zeta _{1}u_{bc}+\zeta _{2}u_{cc},\hspace{8pt} 
\nonumber \\
i\hbar \frac{du_{bc}}{dt} &=&\zeta _{1}^{\dagger }u_{ac},  \nonumber \\
i\hbar \frac{du_{cc}}{dt} &=&\zeta _{2}^{\dagger }u_{ac}.
\end{eqnarray}

Let us take $\Delta _{1}=\Delta _{2}=\Delta =-i\eta $%
\begin{eqnarray}
\zeta _{1} &=&\hbar e^{i\Delta _{1}t}g_{1}a=\hbar \alpha _{1}e^{\eta t}, 
\nonumber \\
\zeta _{2} &=&\hbar e^{i\Delta _{2}t}g_{2}a=\hbar \alpha _{2}e^{\eta t}.
\end{eqnarray}%
Notice that all the three systems of differential equations above are of the
form 
\begin{eqnarray}
i\frac{dx}{dt} &=&\alpha _{1}e^{\eta t}y+\alpha _{2}e^{\eta t}z,  \nonumber
\\
ie^{\eta t}\frac{dy}{dt} &=&\alpha _{1}^{\dagger }x,  \nonumber \\
ie^{\eta t}\frac{dz}{dt} &=&\alpha _{2}^{\dagger }x.
\end{eqnarray}%
We can solve the systems of differential equations using, for instance,
Laplace transformation and we have for the degenerate case 
\begin{eqnarray}
u_{aa}(t) &=&\frac{e^{i\frac{\Delta t}{2}}}{\sqrt{\mu }}\left[ \sqrt{\mu }%
\cos \sqrt{\mu }t-i\frac{\Delta }{2}\sin \sqrt{\mu }t\right] ,  \nonumber \\
u_{ab}(t) &=&-i\frac{e^{i\frac{\Delta t}{2}}}{\sqrt{\mu }}\sin \sqrt{\mu }%
t\alpha _{1},  \nonumber \\
u_{ac}(t) &=&-i\frac{e^{i\frac{\Delta t}{2}}}{\sqrt{\mu }}\sin \sqrt{\mu }%
t\alpha _{2},  \nonumber \\
u_{ba}(t) &=&-i\alpha _{1}^{\dagger }\frac{e^{-i\frac{\Delta t}{2}}}{\sqrt{%
\mu }}\sin \sqrt{\mu }t,  \nonumber \\
u_{bb}(t) &=&1+\alpha _{1}^{\dagger }\alpha _{1}\frac{1}{\sqrt{\nu }}\frac{1%
}{\alpha _{1}^{\dagger }\alpha _{1}+\alpha _{2}^{\dagger }\alpha _{2}}\left[
e^{-i\frac{\Delta }{2}t}\left( i\frac{\Delta }{2}\sin \sqrt{\nu }t+\sqrt{\nu 
}\cos \sqrt{\nu }t\right) -\sqrt{\nu }\right] ,  \nonumber \\
u_{bc}(t) &=&\alpha _{1}^{\dagger }\alpha _{2}\frac{1}{\alpha _{1}^{\dagger
}\alpha _{1}+\alpha _{2}^{\dagger }\alpha _{2}}\left[ e^{-i\frac{\Delta }{2}%
t}\left( i\frac{\Delta }{2\sqrt{\nu }}\sin \sqrt{\nu }t+\cos \sqrt{\nu }%
t\right) -1\right] ,  \nonumber \\
u_{ca}(t) &=&-i\alpha _{2}^{\dagger }\frac{e^{-i\frac{\Delta t}{2}}}{\sqrt{%
\mu }}\sin \sqrt{\mu }t,  \nonumber \\
u_{cb}(t) &=&\alpha _{2}^{\dagger }\alpha _{1}\frac{1}{\alpha _{1}^{\dagger
}\alpha _{1}+\alpha _{2}^{\dagger }\alpha _{2}}\left[ e^{-i\frac{\Delta }{2}%
t}\left( i\frac{\Delta }{2\sqrt{\nu }}\sin \sqrt{\nu }t+\cos \sqrt{\nu }%
t\right) -1\right] ,  \nonumber \\
u_{cc}(t) &=&1+\alpha _{2}^{\dagger }\alpha _{2}\frac{1}{\sqrt{\nu }}\frac{1%
}{\alpha _{1}^{\dagger }\alpha _{1}+\alpha _{2}^{\dagger }\alpha _{2}}\left[
e^{-i\frac{\Delta }{2}t}\left( i\frac{\Delta }{2}\sin \sqrt{\nu }t+\sqrt{\nu 
}\cos \sqrt{\nu }t\right) -\sqrt{\nu }\right] ,
\end{eqnarray}%
where 
\begin{eqnarray}
\mu &=&\frac{\Delta ^{2}}{4}+\alpha _{1}\alpha _{1}^{\dagger }+\alpha
_{2}\alpha _{2}^{\dagger },  \nonumber \\
\nu &=&\frac{\Delta ^{2}}{4}+\alpha _{1}^{\dagger }\alpha _{1}+\alpha
_{2}^{\dagger }\alpha _{2}.
\end{eqnarray}

It is easy to show that for the non-degenerate case, i.e., 
\begin{eqnarray}
\alpha _{1} &=&g_{1}a_{1},  \nonumber \\
\alpha _{2} &=&g_{2}a_{2},
\end{eqnarray}%
we obtain 
\begin{eqnarray}
u_{aa}(t) &=&\frac{e^{i\frac{\Delta t}{2}}}{\sqrt{\mu }}\left[ \sqrt{\mu }%
\cos \sqrt{\mu }t-i\frac{\Delta }{2}\sin \sqrt{\mu }t\right] ,  \nonumber \\
u_{ab}(t) &=&-i\frac{e^{i\frac{\Delta t}{2}}}{\sqrt{\mu }}\sin \sqrt{\mu }%
t\alpha _{1},  \nonumber \\
u_{ac}(t) &=&-i\frac{e^{i\frac{\Delta t}{2}}}{\sqrt{\mu }}\sin \sqrt{\mu }%
t\alpha _{2},  \nonumber \\
u_{ba}(t) &=&-i\alpha _{1}^{\dagger }\frac{e^{-i\frac{\Delta t}{2}}}{\sqrt{%
\mu }}\sin \sqrt{\mu }t,  \nonumber \\
u_{bb}(t) &=&1+\alpha _{1}^{\dagger }\alpha _{1}\frac{1}{\sqrt{\nu _{1}}}%
\frac{1}{\alpha _{1}^{\dagger }\alpha _{1}+\alpha _{2}\alpha _{2}^{\dagger }}%
\left[ e^{-i\frac{\Delta }{2}t}\left( i\frac{\Delta }{2}\sin \sqrt{\nu _{1}}%
t+\sqrt{\nu _{1}}\cos \sqrt{\nu _{1}}t\right) -\sqrt{\nu _{1}}\right] , 
\nonumber \\
u_{bc}(t) &=&\alpha _{1}^{\dagger }\alpha _{2}\frac{1}{\alpha _{1}\alpha
_{1}^{\dagger }+\alpha _{2}^{\dagger }\alpha _{2}}\left[ e^{-i\frac{\Delta }{%
2}t}\left( i\frac{\Delta }{2\sqrt{\nu _{1}}}\sin \sqrt{\nu _{1}}t+\cos \sqrt{%
\nu _{1}}t\right) -1\right] ,  \nonumber \\
u_{ca}(t) &=&-i\alpha _{2}^{\dagger }\frac{e^{-i\frac{\Delta t}{2}}}{\sqrt{%
\mu }}\sin \sqrt{\mu }t,  \nonumber \\
u_{cb}(t) &=&\alpha _{2}^{\dagger }\alpha _{1}\frac{1}{\alpha _{1}^{\dagger
}\alpha _{1}+\alpha _{2}\alpha _{2}^{\dagger }}\left[ e^{-i\frac{\Delta }{2}%
t}\left( i\frac{\Delta }{2\sqrt{\nu _{1}}}\sin \sqrt{\nu _{1}}t+\cos \sqrt{%
\nu _{1}}t\right) -1\right] ,  \nonumber \\
u_{cc}(t) &=&1+\alpha _{2}^{\dagger }\alpha _{2}\frac{1}{\sqrt{\nu _{2}}}%
\frac{1}{\alpha _{1}\alpha _{1}^{\dagger }+\alpha _{2}^{\dagger }\alpha _{2}}%
\left[ e^{-i\frac{\Delta }{2}t}\left( i\frac{\Delta }{2}\sin \sqrt{\nu _{2}}%
t+\sqrt{\nu _{2}}\cos \sqrt{\nu _{2}}t\right) -\sqrt{\nu _{2}}\right] ,
\label{nondeg}
\end{eqnarray}%
where 
\begin{eqnarray}
\mu &=&\frac{\Delta ^{2}}{4}+\alpha _{1}\alpha _{1}^{\dagger }+\alpha
_{2}\alpha _{2}^{\dagger },  \nonumber \\
\nu _{1} &=&\frac{\Delta ^{2}}{4}+\alpha _{1}^{\dagger }\alpha _{1}+\alpha
_{2}\alpha _{2}^{\dagger },  \nonumber \\
\nu _{2} &=&\frac{\Delta ^{2}}{4}+\alpha _{1}\alpha _{1}^{\dagger }+\alpha
_{2}^{\dagger }\alpha _{2}.
\end{eqnarray}

Returning to the degenerate case, in the large detuning limit, we have 
\begin{eqnarray}
\sqrt{\mu } &\approx &\frac{\Delta ^{2}}{2}+\frac{\alpha _{1}\alpha
_{1}^{\dagger }+\alpha _{2}\alpha _{2}^{\dagger }}{\Delta }=\frac{\Delta }{2}%
+\frac{\mid g_{1}\mid ^{2}+\mid g_{2}\mid ^{2}}{\Delta }aa^{\dagger }, 
\nonumber \\
\sqrt{\nu } &\approx &\frac{\Delta ^{2}}{2}+\frac{\alpha _{1}^{\dagger
}\alpha _{1}+\alpha _{2}^{\dagger }\alpha _{2}}{\Delta }=\frac{\Delta }{2}+%
\frac{\mid g_{1}\mid ^{2}+\mid g_{2}\mid ^{2}}{\Delta }a^{\dagger }a,
\end{eqnarray}%
and we get 
\begin{eqnarray}
u_{aa}(t) &=&\exp \left( i\frac{\mid g_{1}\mid ^{2}+\mid g_{2}\mid ^{2}}{%
\Delta }taa^{\dagger }\right) ,  \nonumber \\
u_{ab}(t) &=&0,  \nonumber \\
u_{ac}(t) &=&0,  \nonumber \\
u_{ba}(t) &=&0,  \nonumber \\
u_{bb}(t) &=&1+\frac{\mid g_{1}\mid ^{2}}{\mid g_{1}\mid ^{2}+\mid g_{2}\mid
^{2}}\left[ \exp \left( i\frac{\mid g_{1}\mid ^{2}+\mid g_{2}\mid ^{2}}{%
\Delta }ta^{\dagger }a\right) -1\right] ,  \nonumber \\
u_{bc}(t) &=&\frac{g_{1}^{\ast }g_{2}}{\mid g_{1}\mid ^{2}+\mid g_{2}\mid
^{2}}\left[ \exp \left( i\frac{\mid g_{1}\mid ^{2}+\mid g_{2}\mid ^{2}}{%
\Delta }ta^{\dagger }a\right) -1\right] ,  \nonumber \\
u_{ca}(t) &=&0,  \nonumber \\
u_{cb}(t) &=&\frac{g_{2}^{\ast }g_{1}}{\mid g_{1}\mid ^{2}+\mid g_{2}\mid
^{2}}\left[ \exp \left( i\frac{\mid g_{1}\mid ^{2}+\mid g_{2}\mid ^{2}}{%
\Delta }ta^{\dagger }a\right) -1\right] ,  \nonumber \\
u_{cc}(t) &=&1+\frac{\mid g_{2}\mid ^{2}}{\mid g_{1}\mid ^{2}+\mid g_{2}\mid
^{2}}\left[ \exp \left( i\frac{\mid g_{1}\mid ^{2}+\mid g_{2}\mid ^{2}}{%
\Delta }ta^{\dagger }a\right) -1\right] .
\end{eqnarray}%
If 
\begin{eqnarray}
g_{1} &=&ge^{i\varphi _{1}},  \nonumber \\
g_{2} &=&ge^{i\varphi _{2}},
\end{eqnarray}%
we finally have 
\begin{eqnarray}
u_{aa}(t) &=&\exp \left( i\frac{2g^{2}t}{\Delta }\right) \exp \left( i\frac{%
2g^{2}t}{\Delta }a^{\dagger }a\right) ,  \nonumber \\
u_{ab}(t) &=&0,  \nonumber \\
u_{ac}(t) &=&0,  \nonumber \\
u_{ba}(t) &=&0,  \nonumber \\
u_{bb}(t) &=&\frac{1}{2}\left[ \exp \left( i\frac{2g^{2}t}{\Delta }%
a^{\dagger }a\right) +1\right] ,  \nonumber \\
u_{bc}(t) &=&\frac{e^{i(\varphi _{1}-\varphi _{2})}}{2}\left[ \exp \left( i%
\frac{2g^{2}t}{\Delta }a^{\dagger }a\right) -1\right] ,  \nonumber \\
u_{ca}(t) &=&0,  \nonumber \\
u_{cb}(t) &=&\frac{e^{-i(\varphi _{1}-\varphi _{2})}}{2}\left[ \exp \left( i%
\frac{2g^{2}t}{\Delta }a^{\dagger }a\right) -1\right] ,  \nonumber \\
u_{cc}(t) &=&\frac{1}{2}\left[ \exp \left( i\frac{2g^{2}t}{\Delta }%
a^{\dagger }a\right) +1\right] ,
\end{eqnarray}%
which agrees with the result obtained in \cite{Knight}.\newline

\textbf{Figure Captions} \newline

\textbf{Fig. 1-} Energy states scheme of a three-level atom where $|e\rangle 
$ is the upper state with atomic frequency $\omega _{e}$, $\ |f\rangle $ is
the intermediate state with atomic frequency $\omega _{f}$, $|g\rangle $ is
the lower state with atomic frequency $\omega _{g}$ and $\omega $ is the
cavity field frequency and $\Delta =(\omega _{e}-\omega _{f})-\omega $ is
the detuning. The transition $\mid f\rangle \rightleftharpoons \mid e\rangle 
$ is far enough of resonance with the cavity central frequency such that
only virtual transitions occur between these levels (only these states
interact with field in cavity $C$). In addition we assume that the
transition $\mid e\rangle \rightleftharpoons \mid g\rangle $ is highly
detuned from the cavity frequency so that there will be no coupling with the
cavity field in $C$.\newline

\textbf{Fig. 2- }Set-up for teleportation process. (a) Preparation of a\
Bell state: we send atom $A1$ through a Ramsey cavity $R1$, cavity $C$
prepared initially in a coherent state $(|0\rangle +|1\rangle )/\sqrt{2}$
and through a \ Ramsey cavity $R2$ and atom $A2$ through a Ramsey cavity $R3$%
, cavity $C$ and through a\ Ramsey cavity $R4$. Then we send a two-level
atom $A3$ resonant with the cavity through $C$ in the lower state $%
|f_{3}\rangle $ and through \ Ramsey cavity $R5$ and detect the upper state $%
|e_{3}\rangle $ or $|f_{3}\rangle $ in $D3$. (b) Alice and Bob meet and
generate a Bell state involving atoms $A1$ and $A2$. Then they separate and
Alice keeps atom $A2$ with her and Bob keeps atom $A1$ with him. Later on
Alice decides to teleport an unknown state prepared in atom $A4$ to Bob.
Alice sends atoms $A2 $ and $A4$ through a cavity $C$ prepared initially in
a state $(|0\rangle +|1\rangle )/\sqrt{2}$. After atoms $A2$ and $A4$ have
flown through $C$ she sends a two-level atom $A5$ resonant with the cavity
through $C$ in the lower state $|f_{5}\rangle $ and \ through a Ramsey
cavity $R6$ and then detects $|e_{5}\rangle $ or $|f_{5}\rangle $ in $D5$.
Then she must perform a measurement of the remaining Bell states of the Bell
basis. For this purpose she sends atom $A2$ through the Ramsey cavity $K_{2}$
and $A4$ through Ramsey cavity $K_{4}$. Then, she calls Bob and informs him
the result of her atomic detections in detectors $D2$ and $D4$. Depending on
the results of the Alice's atomic detections and which atomic state of $A5$
she detected, Bob has or not to perform an extra rotation in the Ramsey
cavity $R7$ on the states of his atom $A1.$\newline

\textbf{Fig. 3-} Energy level scheme of the three-level lambda atom where $%
|a\rangle $ is the upper state with atomic frequency $\omega _{a}$, $%
|b\rangle $ \ and $|c\rangle $ are the lower states with atomic frequency $%
\omega _{b}$ and $\omega _{c}$, $\omega $ is the cavity field frequency and $%
\Delta =\omega _{a}-\omega _{b}-\omega =\omega _{a}-\omega _{c}-\omega $ is
the detuning.\newline

\textbf{Fig. 4- }Set-up for teleportation process. (a) Preparation of a\
Bell state: we send atoms $A1$ and $A2$ through a cavity $C$ prepared
initially in a coherent state $(|0\rangle +|1\rangle )/\sqrt{2}$. Then we
send a two-level atom $A3$ resonant with the cavity through $C$ in the lower
state $|f_{3}\rangle $ and through \ a Ramsey cavity $R1$ and detect the
upper state $|e_{3}\rangle $ or $|f_{3}\rangle $ in $D3$. (b) Alice and Bob
meet and generate an Bell state involving atoms $A2$ and $A4$. Then they
separate and Alice keeps atom $A2$ with her and Bob keeps atom $A1$ with
him. Later on Alice decides to teleport an unknown state prepared in atom $%
A4 $ to Bob. She sends atoms $A2$ and $A4$ through a cavity $C$ prepared
initially in a state $(|0\rangle +|1\rangle )/\sqrt{2}$. After atoms $A2$
and $A4$ have flown through $C$ Alice sends a two-level atom $A3$ prepared
initially in the lower state $\mid f_{3}\rangle $ \ and resonant with the
cavity through $C,$ through a Ramnsey cavity $R2$, and detects $\mid
e_{3}\rangle $ or $\mid f_{3}\rangle $ in $D3$. Then she detects the states
of atoms $A2$ and $A4$ in detectors $D2$ and $D4$ and calls Bob and inform
him the result of her atomic detections. Depending on the results of Alice's
atomic detections Bob has or not to perform an extra rotation in the Ramsey
cavity $R3$ on the states of his atom $A4.$\newline

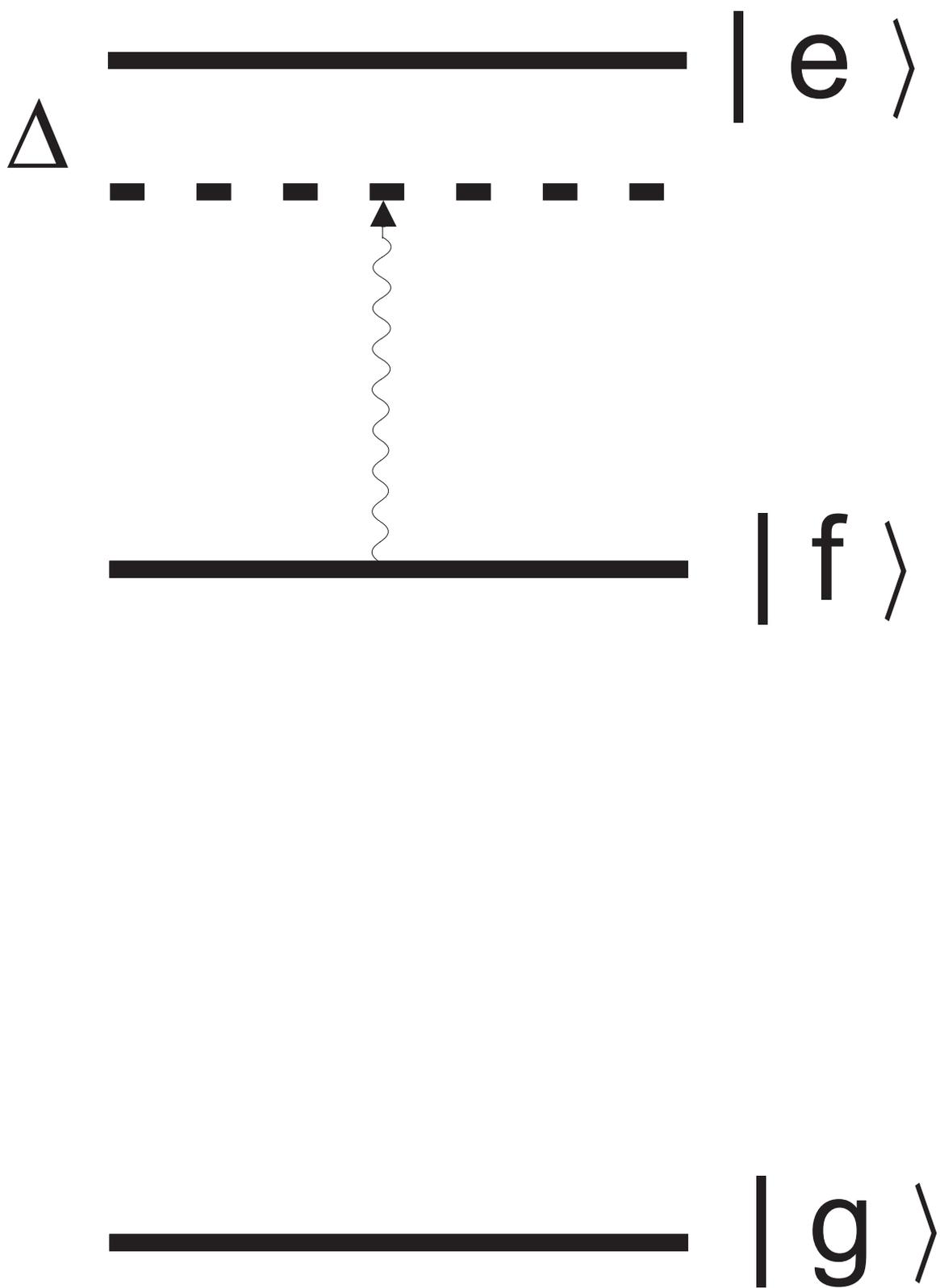
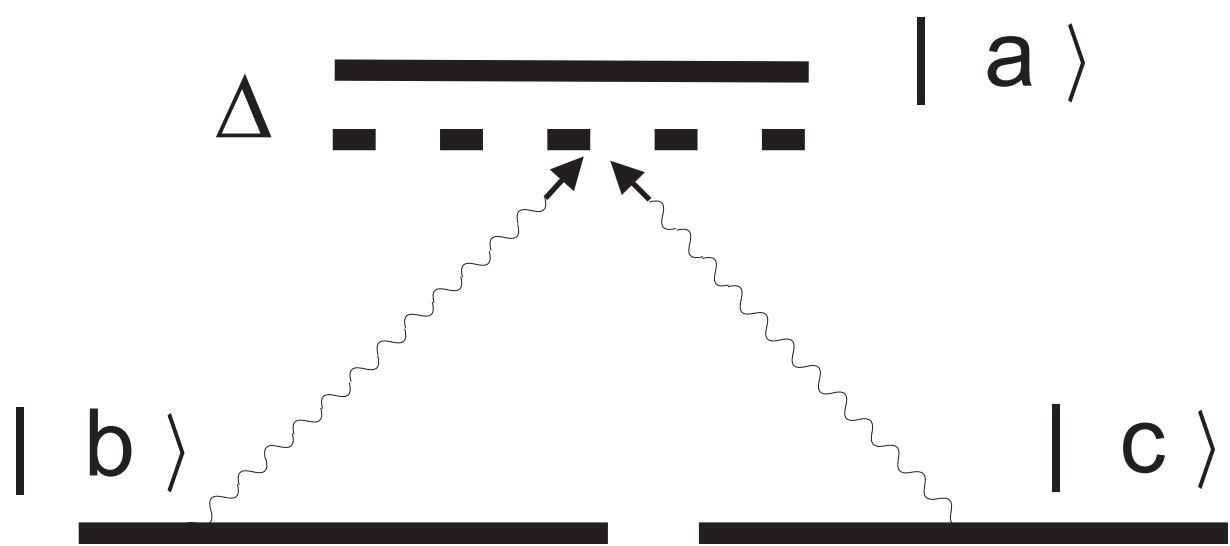
\end{document}